\documentclass[12pt,twoside]{article}
\usepackage{amsmath,amsfonts,amssymb}
\usepackage{latexsym}
\usepackage{dcolumn}
\usepackage{graphicx,epsfig}
\usepackage{amsthm}
\usepackage{hyperref}

\begin{document}

\begin{flushright}
{\tt IISER(Kolkata)/GR-QC\\ \today}
\end{flushright}
\vspace{1.5cm}

\begin{center}
{\Large \bf Quantum Black Hole and the Modified Uncertainty Principle }
\vglue 0.5cm
Barun Majumder \footnote{barunbasanta@iiserkol.ac.in}
\vglue 0.6cm
{\small {\it Department of Physical Sciences,\\Indian Institute of Science Education and Research (Kolkata),\\
Mohanpur, Nadia, West Bengal, Pin 741252,\\India}}
\end{center}
\vspace{.1cm}

\begin{abstract} 
Recently Ali et al. (2009) proposed a Generalized Uncertainty Principle (or GUP) with a linear term in momentum (accompanied by Plank length). Inspired
by this idea we examine the Wheeler-DeWitt equation for a Schwarzschild black hole with a modified Heisenberg algebra which has a linear term
in momentum. We found that the leading contribution to mass comes from the square root of the quantum number $n$ which coincides with Bekenstein's
proposal. We also found that the mass of the black hole is directly proportional to the quantum number $n$ when quantum gravity effects are taken
into consideration via the modified uncertainty relation but it reduces the value of mass for a particular value of the quantum number. 
\vspace{5mm}\newline Keywords: quantum Schwarzschild black hole, Wheeler deWitt quantization, uncertainty principle
\end{abstract}
\vspace{.3cm}


A remarkable amount of research on the physics of black holes over the last few decades has helped us understanding some very fundamental and
unanswered questions of principle. One of them is ``Is the Hawking radiation \cite{i1} truly thermal?''. The most important ingredient to answer this
question involve a consistent quantization of the black hole geometry. Bekenstein first proposed the quantization of black holes \cite{i2}. Later
Vaz et al. \cite{i3} examined the Wheeler-deWitt equation for a static, eternal Schwarzschild black hole and obtained its energy eigenstates
of definite parity. Mass quantization of black holes in terms of the quantum number $n$ implies that Hawking radiation is \textit{not} thermal.
\par 
The idea that the uncertainty principle could be affected by gravity was first given by Mead \cite{i4}. Later modified commutation relations between position and momenta commonly known as Generalized Uncertainty Principle ( or GUP ) were given by candidate theories of quantum gravity ( String Theory, Doubly Special
Relativity ( or DSR ) Theory and Black Hole Physics ) with the prediction of a minimum measurable length \cite{G,i5,i6,i7}. Similar kind of
commutation relation can also be found in the context of Polymer Quantization in terms of Polymer Mass Scale \cite{i8}.
There has been much attention devoted to resolving the quantum corrections to the black hole entropy with the generalized uncertainty principle. Many researchers have expressed a vested interest in fixing the coefficient of the subleading logarithmic term. Using the generalized uncertainty principle as the primary
input, a perturbative calculation of the quantum-corrected entropy, which can readily be extended to any desired order, can be done \cite{i9}.
\par
In this Letter we examine the Wheeler-DeWitt equation for a Schwarzschild black hole with a modified Heisenberg algebra which has a linear term
in momentum. We found that the leading contribution to mass comes from the square root of the quantum number $n$ which coincides with Bekenstein's
proposal. We also found that the mass of the black hole is directly proportional to the quantum number $n$ when quantum gravity effects are taken
into consideration via the modified uncertainty relation but it reduces the value of mass for a particular value of the quantum number. Bina et al.
\cite{b} used the deformed Heisenberg algebra as introduced by \cite{i6} and found similar kind of dependence of mass on the quantum number but
there the mass of the black hole increases with the proportionality factor $n$.
\par
We start with the classical Hamiltonian for a Schwarzschild black hole which can be written as \cite{b34}
\begin{equation}
\label{eq1}
H = \frac{p^2}{2a} + \frac{a}{2} ~,
\end{equation}
where $p_a$ is momentum canonically conjugate to $a$. The phase space co-ordinates $a$ and $p_a$ are deduced from the phase space co-ordinates
$m$ and $p_m$ by means of the canonical transformation $m(t):=M(t,r)$ and $p_m(t):= \int_{-\infty}^\infty \, p_M(t,r)$. The variable $m$ can be
related to the mass $M$ of the black hole when the Einstein's equations are satisfied. To get equation (\ref{eq1}) we have to consider a Hamiltonian quantum theory of a spherically symmetric spacetime. The physical phase space of such a spacetime is spanned by the mass together with the corresponding canonically conjugate momentum. In this phase space we can perform canonical transformations such that the resulting variables describe the dynamical properties of the Schwarzschild black holes. The Wheeler-deWitt equation for the Schwarzschild black hole
can be written as \cite{b34,b35}
\begin{equation}
a^{-s-1} ~\frac{\partial}{\partial a} \big(~a^s ~\frac{\partial}{\partial a} ~\psi (a)~\big) = (a-2M)~\psi (a) ~~,
\end{equation}  
where the substitution $p_a \rightarrow -i \frac{\partial}{\partial a}$ is made. Here we have considered the choice of units where $c=\hbar=G=1$ and
$s$ is the factor ordering parameter. With a particular choice $s=2$ and identifying $R_s=2M$ we get
\begin{equation}
\label{e3}
\frac{1}{a}~\frac{\partial^2 \psi}{\partial a^2} + \frac{2}{a^2}~\frac{\partial \psi}{\partial a} = (a-R_s)~\psi ~~.
\end{equation}
The following transformation $\psi(a)=\frac{U}{a}$ with $x=a-R_s/2$ transforms equation (\ref{e3}) into
\begin{equation}
\label{e4}
-\frac{\partial^2 U}{\partial x^2} + x^2 U = \frac{R_s^2}{4}~U~~.
\end{equation}
The variable $x$ describes the gravitational degrees of freedom. As the total energy of the black hole is included and
the ADM energy is equal to zero so we consider that the energy of excitation of $a$ is not positive. Equation (\ref{e4}) is
exactly the differential equation for a quantum harmonic oscillator with energy levels
\begin{equation}
\frac{R_s^2}{4} = (2n+1)~~,
\end{equation}
$n$ being an non-negative integer. So we get the mass of the black hole as 
\begin{equation}
M^2(n) = 2 (n + \frac{1}{2})~~.
\end{equation}
This result coincides with Bekenstein's argument \cite{i2} as we can see that the mass of the black hole is proportional to $\sqrt{n}$.
\par
The authors in \cite{my5} proposed a generalized uncertainty principle which is consistent with DSR theory, String theory and Black Hole Physics and which says
\begin{equation}
\left[x_i,x_j\right] = \left[p_i,p_j\right] = 0 ,
\end{equation}
\begin{equation}
\label{e8}
[x_i, p_j] = i \hbar \left[  \delta_{ij} -  l  \left( p \delta_{ij} +
\frac{p_i p_j}{p} \right) + l^2  \left( p^2 \delta_{ij}  + 3 p_{i} p_{j} \right)  \right],
\end{equation}
\begin{equation}
\label{e9}
 \Delta x \Delta p \geq \frac{\hbar}{2} \left[ 1 - 2 l \langle p \rangle + 4 l^2 \langle p^2 \rangle \right]~~,
\end{equation}
where $ l=\frac{l_0 l_{pl}}{\hbar} $. Here $ l_{pl} $ is the Plank length ($ \approx 10^{-35} m $). It is normally assumed that the dimensionless
parameter $l_0$ is of the order unity. If this is the case then the $l$ dependent terms are only important at or near the Plank
regime. But here we expect the existence of a new intermediate physical length scale of the order of $l \hbar = l_0 l_{pl}$. We also note
that this unobserved length scale cannot exceed the electroweak length scale \cite{my5} which implies $l_0 \leq 10^{17}$. These equations are
approximately covariant under DSR transformations but not Lorentz covariant \cite{i7}. These equations also imply
\begin{equation}
\Delta x \geq \left(\Delta x \right)_{min} \approx l_0\,l_{pl}
\end{equation}
and
\begin{equation}
\Delta p \leq \left(\Delta p \right)_{max} \approx \frac{M_{pl}c}{l_0}
\end{equation}
where $ M_{pl} $ is the Plank mass and $c$ is the velocity of light in vacuum. The effect of this proposed GUP is well studied recently for some well known
physical systems in \cite{my5,b8,zz}.
\par
Equations (\ref{e8}) and (\ref{e9}) represents modified Heisenberg algebra. But the interesting part of these two relations is the term which is linear
in $l$ ($= l_0l_{pl}$) with $p$. In the next section of this Letter we will study the effect of this linear term in Plank length in the context of
Wheeler-deWitt quantization of a Schwarzschild black hole. Our analysis will be perturbative as in the first approximation we will neglect
terms ${\cal O}(l^2)$ and more. Inspired by this idea, for our purpose we will consider the modified Heisenberg algebra (modified uncertainty principle)
where $\bf{x}$ and $\bf{p}$ obeys the relation ($\alpha >0$)
\begin{equation}
\label{e12}
[\mathbf{x}~,~\mathbf{p}] = i~(1-\alpha~\mathbf{p})~~.
\end{equation}
We have used units with $\hbar=1$. We can see that if $\alpha=2l$ this is the same relation as that of equation (\ref{e8}) only upto a linear
term in $l$. It can be shown that the smallest uncertainty in position occurs when $\langle \bf{p} \rangle$ $=0$ and $\Delta x_{min}=\frac{\alpha}{2}$. The
momentum space wave function can be written as $\psi(p)=\langle p~|\psi \rangle $. On a dense domain in Hilbert space $\bf{x}$ and $\bf{p}$ act
as operators such that
\begin{equation}
\mathbf{p} ~\psi(p) = p ~\psi (p)~~,
\end{equation}
\begin{equation}
\label{e14}
\mathbf{x} ~\psi (p) = i ~\Big[~(1-\alpha ~p)~\frac{\partial}{\partial p}~\Big]~\psi(p)~~.
\end{equation}
This representation respects the commutation relation (\ref{e12}) and the scalar product of two arbitrary wave functions in this representation is given by
\begin{equation}
\langle \phi ~|~ \psi \rangle = \int_{-\infty}^\infty ~dp~ \phi^*(p)~\psi (p)~~.
\end{equation}
Considering the standard derivation of the uncertainty relation we can see that if the state $|\psi \rangle $ obeys 
$\Delta x \Delta p = \frac{|\langle [x,p]\rangle|}{2}$ then it will obey the relation
\begin{equation}
\label{e16}
\Big(~\mathbf{x} - \langle \mathbf{x}\rangle + \frac{\langle [\mathbf{x},\mathbf{p}]\rangle}{2(\Delta p)^2}~(\mathbf{p}~ - \langle \mathbf{p}\rangle)\Big)~ 
|\psi \rangle = 0 ~~.
\end{equation}
The states of absolutely maximal localization can only be obtained for $\langle \mathbf{p}\rangle = 0$ with critical momentum uncertainty
$\Delta = \frac{2}{\alpha}$. With equation (\ref{e14}) and (\ref{e16}) we can calculate these states in momentum space and the states are
\begin{equation}
\psi_{\langle x \rangle} (p) = {\cal N} ~(1-\alpha ~ p)^{i\frac{\langle x \rangle}{\alpha}}~ e^{-\frac{p}{4}}~(1-\alpha ~ p)^{-\frac{1}{4\alpha}}~~.
\end{equation}
Normalization of this wave function cannot be done as the integral required for this diverges. $(1-\alpha ~ p)$ contains the first two terms of the
series form of $e^{-\alpha p}$. As we have mentioned earlier that our approach is in some sense perturbative, so here we use an approximation
$(1-\alpha ~ p) \approx e^{-\alpha p}$. Using this we get
\begin{equation}
\psi_{\langle x \rangle} (p) = {\cal N}~ e^{i\langle x \rangle p} ~~.
\end{equation}
Now we can use a delta-function normalization (for example \cite{Bransden}) and get 
\begin{equation}
\psi_{\langle x \rangle} (p) = \frac{1}{\sqrt{2\pi}}~ e^{i\langle x \rangle p} ~~,
\end{equation}
where we have used the relation
\begin{equation}
\langle \phi_{\langle x' \rangle}|~\psi_{\langle x \rangle}\rangle = \int_{-\infty}^{\infty}~\phi_{\langle x' \rangle}^*(p)~\psi_{\langle x \rangle}(p)~dp~
= \delta (\langle x' \rangle - \langle x \rangle) ~~.
\end{equation}
We could have also used the idea of box normalization for our purpose. The maximal localization states for a deformed Heisenberg algebra with a
linear term in $p$ in the commutation relation is a serious issue because the normalization is not possible. Kempf et al. \cite{i6} first
made use of the deformed algebra
\begin{equation}
[\mathbf{x}~,~\mathbf{p}] = i~(1+\sigma~\mathbf{p}^2)~~~~~~~~~~~,~~\sigma >0
\end{equation}
where $\sigma $ is the deformation parameter. This relation is also seen in perturbative string theory \cite{G}. Here the normalization
of the maximal localization states can be easily done as the integral $ \int_{-\infty}^{\infty}\frac{dp}{1+\sigma p^2} $ converges.
\par
It is a well established fact that the generalized uncertainty relation incorporates a smallest possible length in quantum mechanics which is an effective consequence of any quantum theory of gravity. This uncertainty relation especially refers to the variables of quantum mechanics (position and
momentum of a particle). Here we can extend this idea for the quantization of the gravitational degree of freedom and its canonically conjugate momentum
which leads to the Wheeler-deWitt equation. In principle, we are making an assumption that the variables describing the modified uncertainty relation
given by equation (\ref{e12}) can be treated as variables describing the gravitational degrees of freedom of the Schwarzschild black hole. 
We now apply the above mentioned modified uncertainty relation to study a quantum Schwarzschild black hole. If we apply equation (\ref{e14}) to
equation (\ref{e4}) we get the modified Wheeler-deWitt equation for a Schwarzschild black hole and it is written as
\begin{equation}
\label{e22}
(1-\alpha p)^2~\frac{\partial^2 \psi(p)}{\partial p^2} - \alpha (1-\alpha p)~\frac{\partial \psi(p)}{\partial p} - (p^2 - M^2)~\psi (p) = 0 ~~.
\end{equation}
A change of variable from $p$ to
\begin{equation}
\eta \equiv \ln (1-\alpha p)
\end{equation}
casts equation (\ref{e22}) in the form 
\begin{equation}
\label{e24}
\frac{\partial^2 \psi}{\partial \eta^2} - \big[ a e^{2\eta} + b e^{\eta} + k^2 \big]~ \psi = 0 ~~,
\end{equation}
where $a=\frac{1}{\alpha^4}$, $b=-\frac{2}{\alpha^4}$ and $ k^2 = \frac{1}{\alpha^4} - \frac{M^2}{\alpha^2}$. Now with a new change of variable
$z \equiv e^{\eta}$ and $ w \equiv z^{-k}~\psi $, we get equation (\ref{e24}) in the form
\begin{equation}
z~\frac{\partial^2 w}{\partial z^2} + (2k+1)~ \frac{\partial w}{\partial z} - (az+b)~w = 0~~.
\end{equation}
Finally with the following transformation 
\begin{equation}
w\equiv e^{-\xi /2}~\mathit{f} (\xi)  ~~~~ \text{and} ~~~~\xi \equiv 2\sqrt{a}~z ~~,
\end{equation}
we can arrive at the equation 
\begin{equation}
\label{e27}
\xi ~ \mathit{f}'' + (2k + 1 - \xi)~\mathit{f}' - \bigg[~\frac{(2k+1)\sqrt{a}+b}{2\sqrt{a}}\bigg]~\mathit{f} = 0 ~~.
\end{equation}
Here prime denotes differentiation with respect to $\xi$. $\mathit{f}(\xi)$ should be non-singular at $\xi=0$ and we require a polynomial
solution to equation (\ref{e27}). This requirement imposes the following condition on the coefficient of $\mathit{f}$ :
\begin{equation}
\label{e28}
\frac{(2~k+1)~\sqrt{a}+b}{2~\sqrt{a}} = - n~~,
\end{equation}
where $n$ is a non-negative integer \cite{bell}. Equation (\ref{e27}) now becomes 
\begin{equation}
\xi ~ \mathit{f}'' + (2k + 1 - \xi)~\mathit{f}' +n~\mathit{f} = 0 ~~.
\end{equation}
The solution is known in terms of associated Laguerre polynomials ${\cal L}_n^{2k}(\xi)$ \cite{bell}. The condition (\ref{e28}) gives
the mass eigenvalues for the black hole. After a straightforward algebra it says
\begin{equation}
M^2(n) = 2~(n+\frac{1}{2}) - \alpha^2 ~\big[n^2 + n + \frac{1}{4}~\big] ~~. 
\end{equation}
This result coincides with Bekenstein's proposal \cite{i2} that mass of a black hole is proportional to $\sqrt{n}$. Our leading
contribution to mass is also from $\sqrt{n}$. If we use the deformed algebra of Kempf et al. \cite{i6} we can conclude that
in quantum gravity regime the mass of a black hole is proportional with the quantum number $n$ \cite{b}. Interestingly we also found
with a perturbative approximation that the mass of a black hole is proportional to $n$ when quantum gravity effects are taken into
consideration. We also notice that the modified uncertainty relation which we have used reduces the value of mass for a
particular $n$. So in this Letter we have studied the quantum Schwarzschild black hole in the modified uncertainty principle
framework. It is quite difficult to get and normalize the maximal localization states in our representation. We perturbatively
found that the leading contribution to the mass eigenvalue comes from $\sqrt{n}$ which is in agreement with Bekenstein's proposal. We also
found that there are contributions from the quantum number $n$. Though the form of the uncertainty relation which we have used is
different from the one used in \cite{b} but our results are similar (mass proportional to $n$). We also noticed the fact that
our modified uncertainty relation lessens the mass of the black hole for a particular value of $n$.


\section*{Acknowledgements}
The author is very much thankful to Prof. Narayan Banerjee for helpful discussions and guidance.

\end{document}